\documentclass[doublecol]{epl2}
\usepackage{graphicx}
\newcommand{\beq}{\begin{equation}}
\newcommand{\eneq}{\end{equation}}

\begin{document}

\title{Realization of a two-channel Kondo model with Josephson junction networks}

\author{Domenico Giuliano \inst{1} \and Pasquale Sodano \inst{2} }

\institute{\inst{1} Dipartimento di Fisica, Universit\`a della Calabria
             and I.N.F.N., Gruppo Collegato di Cosenza, Arcavacata di
             Rende, I-87036, Cosenza, Italy
              \\
  \inst{2}  International Institute of Physics, Universidade
  Federal do Rio Grande do Norte, 59012-970, Natal, Brazil  and
  INFN, Sezione di Perugia, Via A. Pascoli, I-06123, Perugia, Italy}


\abstract{ We show that- in the quantum regime- a Josephson junction rhombi chain
(i.e. a Josephson junction chain made by rhombi formed by joining 4 Josephson junctions)
may be effectively mapped onto a quantum Hamiltonian describing Ising spins in a
transverse magnetic field with open boundary conditions. Then, we elucidate how a
Y-shaped network fabricated with 3 Josephson Junction Rhombi chains may be used as a
quantum device realizing the two channel Kondo model recently proposed by Tsvelik in
\cite{tse2}. We point out that the emergence of a 2 channel Kondo effect in this
superconducting network may be probed through the measurement of a pertinent Josephson current.}

\maketitle

{\it Introduction}
\\
The Kondo effect arises from the  (Kondo) antiferromagnetic coupling between  the spin of magnetic impurities
and of itinerant electrons \cite{hewson}. When the number of ''channels'' of conduction electrons is
equal to two times the spin of the impurity, as the temperature $T$ goes below
the dynamically generated Kondo temperature $T_K$,  the Kondo
coupling leads to the formation of the
''Nozier\`es-Fermi liquid''  state, in which  the spin of
itinerant electrons effectively screens the magnetic impurity,
which is traded by a phase shift $\pi/2$ in the electronic
wavefunctions \cite{hewson,noz}.  A different state is
realized when the number $K$ of channels of itinerant electrons is larger
than 2$S$, with $S$ being the impurity spin: as  the electrons tend to ''over-screen''
the impurity \cite{blandin,zawa}, the residual degeneracy resulting from
over-screening yields   a non-Fermi liquid state \cite{destri,wiegmann}, with
peculiar properties, such as, for instance,  a remarkable power-law
dependence on $T$ of the resistivity  (for instance, for $K=2$ one
finds a dependence   on $(T / T_K)^{\frac{1}{2}}$
\cite{aflud}.)

Despite  the great interest in many-channel Kondo models,
their physical realizations, even in controlled devices and in the simplest
possible case, the 2-channel Kondo (2CK)-model, have been, so far,
extremely difficult \cite{GiuTa1,ggh} to attain, due to the need for a perfect symmetry between the
couplings of the spin density from the two channels to the spin
of the impurity. A neat idea to circumvent this problem has been recently proposed
by Tsvelik in a Y-junction of three one-dimensional quantum Ising models (1QIM)s, joined
at the inner edges of the three chains \cite{tse2}. In this proposal, when the relevant
parameters are pertinently tuned, a Y-junction of quantum Ising chains hosts \cite{tse2}
the two-channel Kondo effect. A similar approach has been used in \cite{andcrampo} yielding a spin 
network realization of the four channel Kondo model. These proposals are particularly attractive as spin models
have been known, since a long time \cite{matsua}, to provide reliable and effective
descriptions of quantum coherent phenomena in condensed matter systems.
As a result one  may hope to probe multi-channel Kondo effects in a variety of controllable, and yet 
robust, experimental settings such as the ones provided by degenerate
Bose gases confined in an optical lattice \cite{greiner,grst}, or  quantum Josephson
junction networks (JJN)s \cite{giuso}.

JJNs  are a quite versatile tool for the quantum engineering of reliable devices since
the fabrication and manipulation techniques so far developed (for a review see,
for instance, Ref.\cite{haviland}) led to a quite
good level of confidence on the accuracy of both fabrication and control parameters.
In addition, JJNs in the quantum regime (i.e., when the junctions used to fabricate the 
network are such that the capacitive energy is much bigger than the Josephson energy) may
 be well described by effective spin models whose
relevant parameters are determined from the knowledge of the fabrication and control
parameters of the JJN \cite{giusonew,giuso1}. Furthermore, a pertinent design of certain
JJNs may facilitate the emergence of two level quantum systems with a high degree of
quantum coherence \cite{giusonew,giuso1,giuso2} and Josephson junction rhombi chains (JJRC) 
\cite{cata,propov} are known to induce local $4e$ superconducting
correlations, corresponding to pairing of Cooper pairs in a tunneling process across a
quantum impurity \cite{giuso_epl,giuso3,doucot}.

In this letter, we show how the Tsvelik's realization of the 2CK-effect may be
implemented in a Y-junction of JJRCs. For this purpose,  we shall first
show that, in an effective description keeping only low-energy, long-wavelength excitations,
a single JJRC in the quantum regime may be mapped onto a quantum Ising chain with open boundaries, 
whose parameters are determined by the fabrication parameters of the superconducting network. Then, we elucidate
how three of such chains may be glued together into a Y-junction allowing for the emergence of 
a 2CK regime whose signature may be detected through the measurement of a pertinent dc-Josephson current.

{\it Realization of a Quantum Ising model using a JJRC.}
\\
Within our JJN-realizazion of the 1QIM, a single spin is realized  with  a circular
4-junction array made by four superconducting grains, each one with a  charging energy $E_C$, biased with
a voltage $V_g$, and coupled to the nearest neighboring grains with Josephson energy $J$.
When $E_C \gg J$ and  $V_g$ is tuned so to make the states with ${\cal N}$ and
${\cal N}+1$ Cooper pairs degenerate with each other,   each grain
may be regarded as a quantum spin-1/2 degree of freedom $\vec{S}_j$, acting within
the subspace spanned by  the two states above  \cite{giuso_epl,giuso3}.   As a result,
a single rhombus is well-described by the
effective spin Hamiltonian $H_C = - J \sum_{ j = 1}^4 \{ e^{- \frac{i}{4} \varphi} S_j^+ S_{j+1}^- + {\rm h.c.} \}
- h \sum_{ j = 1}^4 S_j^z$, with $\varphi$ being  the magnetic flux piercing the rhombus (in units of
quantum of flux $\Phi_0^* = \frac{2e}{hc}$), and  the magnetic field $h (\ll J )$ corresponding to  a
possible slight detuning of $V_g$ off the exact degeneracy value $V_g^*$.
As one sets $\varphi = \pi$, the ground state of $H_C$ becomes twofold degenerate, and
it is spanned by the two states
\begin{eqnarray}
 | \Uparrow \rangle &=&
\frac{1}{2 \sqrt{2}} \{ \sqrt{2} [ |  \uparrow \downarrow \uparrow \downarrow \rangle +
| \downarrow \uparrow \downarrow \uparrow \rangle ] + | \uparrow \uparrow \downarrow \downarrow \rangle
\nonumber \\
&+& | \downarrow \downarrow  \uparrow \uparrow \rangle + | \uparrow \downarrow \downarrow  \uparrow\rangle
+ | \downarrow  \uparrow \uparrow  \downarrow \rangle \}
\end{eqnarray}
\noindent
\noindent
and

\begin{eqnarray}
 | \Downarrow \rangle &=& \frac{1}{2 \sqrt{2}} \{ \sqrt{2} [ |  \uparrow \downarrow \uparrow \downarrow \rangle -
| \downarrow \uparrow \downarrow \uparrow \rangle ] - i  | \uparrow \uparrow \downarrow \downarrow \rangle
\nonumber \\
&-& i | \downarrow \downarrow  \uparrow \uparrow \rangle + i  | \uparrow \downarrow \downarrow  \uparrow\rangle
+ i | \downarrow  \uparrow \uparrow  \downarrow \rangle \}.
\end{eqnarray}
\noindent
The states
$ | \Uparrow \rangle , | \Downarrow \rangle$ are two spin singlets,  separated
from higher-energy states by a gap $\sim J$. Their emergence explicitly manifests
the $Z_2$-degeneracy of the ground state of $H_C$, and ultimately allows, as we shall see
in more detail in the following , not only to associate a collective spin-1/2 variable to each rhombus,
but also to describe the JJRC as a $Z_2$-symmetric Ising chain.
\begin{figure}
\centering \includegraphics*[width=0.75\linewidth]{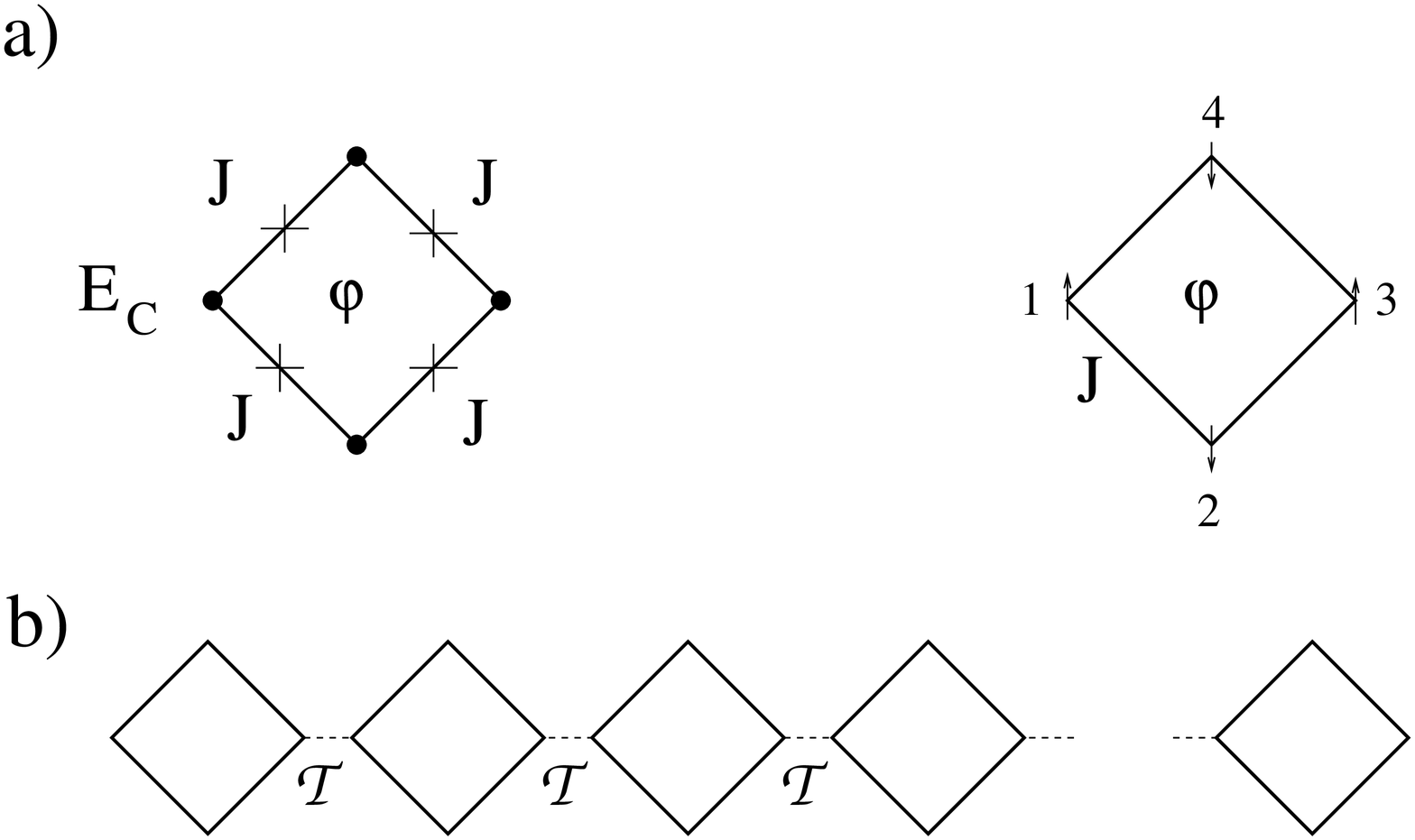}
\caption[]{{\bf a)} the single rhombus: with the parameters chosen as in the
main text,  each grain may be described by a spin-1/2 degree of freedom, interacting with
the nearest-neighboring sites with magnetic exchange constant equal to $J$; \\
{\bf b)} The rhombi chain mapping onto the one-dimensional quantum Ising model. }
\label{f_one}
\end{figure}
\noindent

The JJRC is realized as a chain of $\ell$ rhombi like in Fig.\ref{f_one},
all equal to each other, each one pierced by a magnetic flux $\varphi \sim \pi$. The low-energy effective Hamiltonian
is obtained  by truncating the Hilbert space of the states of each rhombus $_p$ only to
its two groundstates $ | \Uparrow \rangle_p,| \Downarrow \rangle_p$. Accordingly,
the rhombus is described  in terms of a
''collective'' quantum spin operator  ${\bf S}_p = ( S^x_p , S^y_p , S^z_p )$, with
 $p   = 1 , \ldots , \ell$, and $S^a_p =  \frac{1}{2}
\: \sum_{\sigma , \sigma' = \Uparrow , \Downarrow} \: | \sigma \rangle \langle \sigma' | \tau^a_{\sigma , \sigma'}$,
with  $\tau^x , \tau^y , \tau^z$ being the Pauli matrices. To engineer a 1QIM with
the spins  ${\bf S}_p $, we   assume that, say, the grain at site 3 of rhombus $p$
is coupled to the grain at site 1 of rhombus $p+1$, with  Josephson energy  ${\cal T}$
such that  ${\cal T} \ll J$. The corresponding ''microscopic'' Hamiltonian describing such a
chain is given by

\begin{eqnarray}
H_{\rm micro} &=&  - J \sum_{ p = 1}^\ell  \sum_{ j =1}^4 \{ e^{-\frac{i}{4} \varphi} S_{p , j}^+ S_{
p , j+1}^- + {\rm h.c.} \} \\
&-&
h\sum_{p=1}^\ell   \sum_{ j = 1}^4 S_{p , j}^z - {\cal T}   \sum_{ p = 1}^{\ell - 1}
\{ S_{p , 3}^+ S_{p+1 , 1}^- + {\rm h.c.} \}\nonumber
\:,
\label{qi.3}
\end{eqnarray}
\noindent
with the last contribution to the right-hand side of Eq.(\ref{qi.3}),
$H_\tau =  - {\cal T}   \sum_{ p = 1}^{\ell - 1}
\{ S_{p , 3}^+ S_{p+1 , 1}^- + {\rm h.c.} \} \equiv \sum_{p = 1}^{\ell - 1}
H_{p, p +1}$, describing  the Josephson coupling between
nearest-neighboring rhombi. To map  $H_{\rm micro}$ onto a 1QIM-Hamiltonian,
one has to project it onto the low-energy subspace ${\cal F} = \otimes_{p = 1}^\ell \{ {\rm Span} [| \Uparrow\rangle_p ,
 | \Downarrow \rangle_p ] \}$, with  ${\rm Span} [ | \Uparrow \rangle_p ,
 | \Downarrow \rangle_p ]$ being the space spanned by $| \Uparrow \rangle_p ,
 | \Downarrow \rangle_p$. In doing so, one readily sees that, since the term $\propto {\cal T}$ in $H_{\rm micro}$
takes a state originally lying within ${\cal F}$ out of the subspace, the projection
gives 0 to first order in  ${\cal T}$. To recover a nonzero result,  one must necessarily sum over
''virtual'' transitions from and back into  ${\cal F}$. This can be systematically done by performing a second-order
Schrieffer-Wolff (SW)  sum.
The SW-procedure requires building excited states at rhombus $_p$, with eigenvalue of
 $S_{p;T}^z = \sum_{ j = 1}^4 S_{p;j}^z$ equal to $\pm 1$. The eigenstate with
$S_{p;T}^z = 1$ and energy  $\epsilon_1 ( k ) = - 2 J \cos ( k +  \frac{\varphi}{4} ) - 2h$ is
given by $| 1 , k \rangle_p = \frac{1}{2} \sum_{ j = 0}^3 e^{ i k j } | 1 , j \rangle_p$, with
$k = \frac{ 2 \pi r \ell}{4}$ ( $r = 0 , 1 , 2 ,3$), and
$ | 1 , j \rangle_p$ being  the state of rhombus $_p$ with  all the spins
 $\uparrow$, except the one at site $j$. At variance, the eigenstate with
$S_{p;T}^z = - 1$ and energy $ \epsilon_{-1} ( k ) = - 2 J \cos ( k - \frac{\varphi}{4} ) + 2 h$
is given by $ | - 1 , k \rangle_p = \frac{1}{2} \sum_{ j = 0}^3 e^{ i k j } | - 1 , j \rangle_p$, with
$ | - 1 , j \rangle_p$ being  the state of rhombus $_p$ with all the spins
$\downarrow$, except the one at site $j$. Denoting, now, with $| X \rangle_p$ a generic state of rhombus $_p$ with
either $S_{T;p}^z = \pm 1$, the SW procedure allows for writing the
effective Hamiltonian for the system to ${\cal O} ( {\cal T}^2 / J )$
in terms of matrix elements of $H_{p,p+1}$ between states of
the form  $| \sigma \rangle_p \otimes | \rho \rangle_{p+1} $
($\sigma , \rho = \Uparrow , \Downarrow$) and states
 involving the $| X \rangle_p$'s. This yields nontrivial
 matrix elements between  $| \sigma \rangle_p \otimes | \rho \rangle_{p+1} $
 and  $| \sigma' \rangle_p \otimes | \rho' \rangle_{p+1} $,
 defining an effective Hamiltonian $H_{\rm Eff}^{p,p+1}$ such that

\begin{eqnarray}
&& \{ ~_p\langle \sigma' | \otimes ~_{p+1} \langle \rho' | \} H_{\rm Eff}^{p,p+1}
\{ | \sigma \rangle_p \otimes | \rho \rangle_{p+1} \} \nonumber \\
&=& \sum_{X , X'} \Biggl\{
\frac{ \{ ~_p \langle \sigma' | \otimes ~_{p+1} \langle \rho' | \}  H_{p,p+1} \{  | X \rangle_p
\otimes  | X' \rangle_{p+1} \} }{ E_0
 - E_X - E_{X'} } \nonumber \\
 &\times&  \{   ~_p \langle X | \otimes ~_{p+1} \langle X | \}
 H_{p,p+1} \{ | \sigma \rangle_p \otimes | \rho \rangle_{p+1} \} \Biggr\}
\:\:\:\: ,
\label{qi.4}
\end{eqnarray}
\noindent
with $E_0$ being the groundstate energy of $H_C$ and $E_X , E_{X'}$ being
the energies of  $| X \rangle_p $ and of $| X' \rangle_{p+1}$, respectively.
From the explicit result for the nonzero matrix elements of
$H_\tau$, one finds that the matrix elements of $ H_{\rm Eff}^{p,p+1}$
in Eq.(\ref{qi.4}) can be written as a sum of the matrix elements of
two operators, the former one being given by

\beq
H_{\rm Eff; (A)}^{p,p+1}   =   - \frac{{\cal T}^2}{4 J}  \{  {\bf I}_p
{\bf I}_{p+1} - 4 S^x_p   S^x_{p+1} \}
\:\:\:\: ,
\label{efham.2}
\eneq
\noindent
where ${\bf I}_p$ denotes the identity operator acting on the low-energy subspace of
rhombus $p$. The latter operator is instead given by

\begin{eqnarray}
 &&
H_{\rm Eff; (B)}^{p,p+1}     =  \frac{{\cal T}^2}{4 J } \times \nonumber \\
&& \biggl\{ \left[ \frac{1}{4} {\bf I} + \frac{\sqrt{2}}{8} S^z  -
\frac{1}{8} S^x \right]_p   \left[ \frac{1}{4} {\bf I} + \frac{\sqrt{2}}{8} S^z  +
\frac{1}{8} S^x \right]_{p+1} \nonumber \\
&& +  \left[ \frac{1}{4} {\bf I} + \frac{\sqrt{2}}{8} S^z +
\frac{1}{8} S^x \right]_p  \left[ \frac{1}{4} {\bf I} + \frac{\sqrt{2}}{8} S^z-
\frac{1}{8} S^x \right]_{p+1} \nonumber \\
&& - \left[ \frac{1}{4} {\bf I} - \frac{\sqrt{2}}{8} S^z -
\frac{1}{8} S^x \right]_p   \left[ \frac{1}{4} {\bf I} - \frac{\sqrt{2}}{8} S^z +
\frac{1}{8} S^x \right]_{p+1} \nonumber \\
&& - \left[ \frac{1}{4} {\bf I} - \frac{\sqrt{2}}{8} S^z +
\frac{1}{8} S^x \right]_p   \left[ \frac{1}{4} {\bf I} - \frac{\sqrt{2}}{8} S^z -
\frac{1}{8} S^x \right]_{p+1} \biggr\}
\nonumber \\
&& =  \frac{\sqrt{2} {\cal T}^2}{16 J} \{ S_p^z   {\bf I}_{p+1} + {\bf I}_p  S_{p+1}^z \}
\:\:\:\: .
\label{efham.3}
\end{eqnarray}
\noindent
Adding up the Eqs.(\ref{efham.2},\ref{efham.3}), one eventually
finds (besides an irrelevant over-all constant

\beq
H_{\rm Eff; (A)}^{p,p+1} + H_{\rm Eff ; (B) }^{p,p+1}=   J_x S_p^x S_{p+1}^x - H ( S_p^z + S_{p+1}^z )
\;\;\;\; ,
\label{esham.4}
\eneq
\noindent
with $J_x = \frac{{\cal T}^2}{J} $ and $H = - \frac{ \sqrt{2} {\cal T}^2}{16 J}$. On summing over
the index $ p = 1 , \ldots ,\ell$, one finally obtains

\beq
H_{\rm 1QIM} = J_x \sum_{ p = 1}^{\ell - 1}  S_p^x S^x_{p+1} - 2  \sum_{p = 1}^\ell H_p  S_p^z
\:\:\:\: ,
\label{esham.5}
\eneq
\noindent
with $J_x = \frac{ {\cal T}^2}{J}$,  $H_p = H  = - \frac{ \sqrt{2} {\cal T}^2}{16 J}$ for $p = 2 , \ldots , \ell - 1$
and $H_1 = H_\ell = H /2$ two boundary magnetic fields accounting for the chain' s open boundaries. From the explicit formulas for $J_x$ and $H_p$ one sees that,
besides a boundary magnetic fields, which does not affect the bulk
phase diagram, since $J_x > 4 |H|$ by construction , the effective 1QIM describing the JJRC
is in its antiferromagnetic phase, corresponding to the spontaneous breaking of the
spin-parity $Z_2$-symmetry $S_p^x \to - S_p^x$, $S_p^z \to S_p^z$.

\begin{figure}
\centering \includegraphics*[width=1.\linewidth]{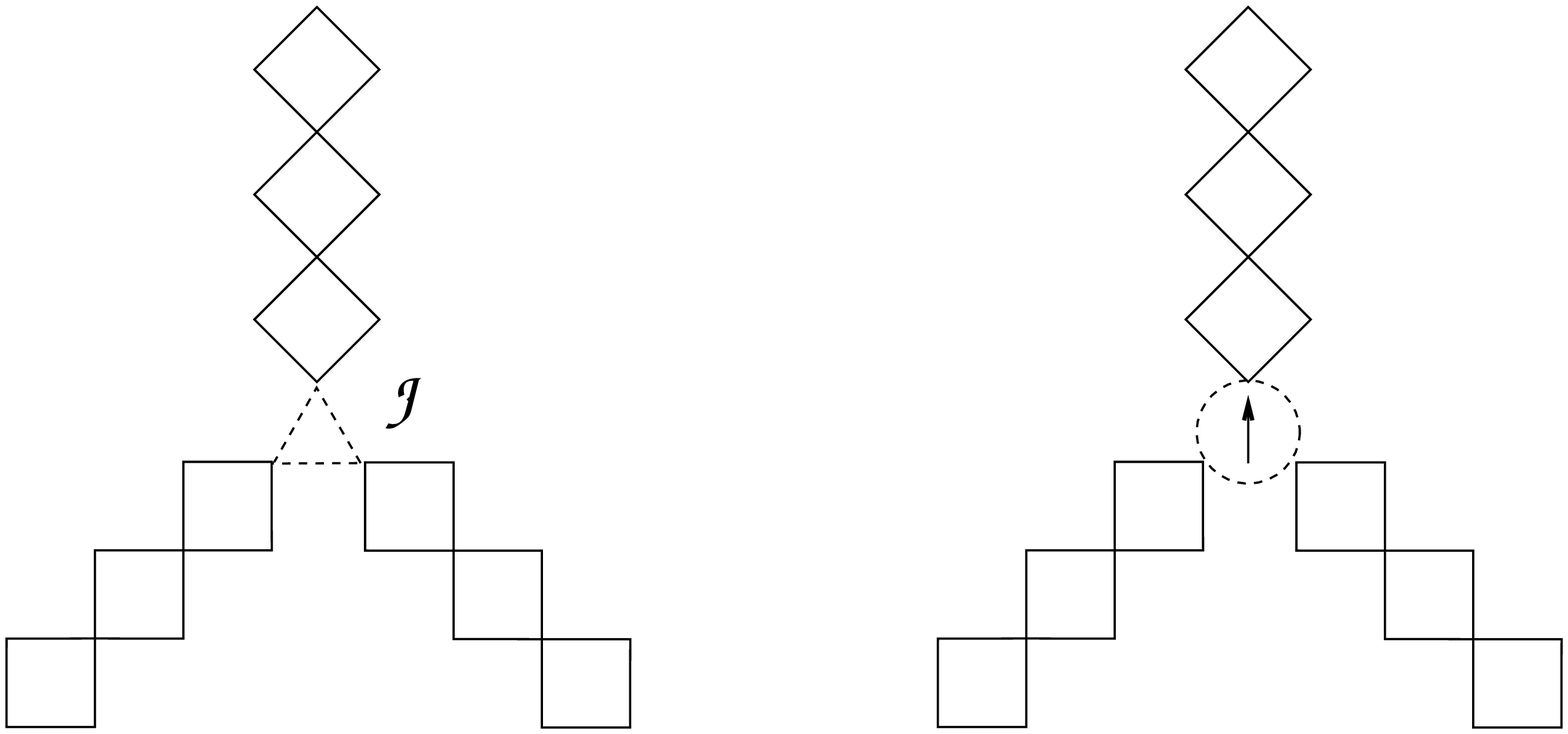}
\caption[]{{\bf a)} $Y$ junction of three rhombi chains: the spins
at the endpoints of the three chains are coupled to each other, with an
effective coupling strength ${\cal J}$; {\bf b)} representation of the central region by an effective spin-1/2 }
\label{y_juna}
\end{figure}

{\it Junction of three JJRCs and mapping onto the 2-channel Kondo model}
\\
To actually show how 2CK-model can be actually realized in a pertinently designed
JJN, we now discuss how to realize Tsvelik's  Y junction  of 1QIM's within
a Josephson junction network.  In order to couple three JJRCs at their endpoints, one needs to
consider the JJN depicted in Fig.\ref{y_juna}, where
the dashed lines correspond to  Josephson couplings between, say, sites number 2 of the endpoint-rhombus of
each chain, with Josephson energy ${\cal J}$, and corresponding ''microscopic'' Hamiltonian given by

\beq
H_{\rm MB;J} = - {\cal J} \{ S_{1,1,2}^+ S_{2,1,2}^- + S_{2,1,2}^+ S_{3,1,2}^- + S_{3,1,2}^+ S_{1,1,2}^- + {\rm h.c.} \}
\:\:\:\: .
\label{yj.1}
\eneq
\noindent
(In Eq.(\ref{yj.1}), the first index of the microscopic spin operator, $\lambda = 1 ,2 ,3 $,
labels the three chains, the second index labels the position of the rhombus ($p=1$ for all
three the chains), the third index labels the position of the single spin within
rhombus $p=1$ of the corresponding chain.) In order, now, to project
$H_{\rm MB;J}$ onto the subspace ${\cal F}$, one may resort to the same SW-procedure
we used to derive the 1QIM-Hamiltonian in Eq.(\ref{esham.5}). As a result,
one eventually trades $H_{\rm MB;J}$ for an effective boundary Hamiltonian $H_B$,
 involving only the spins ${\bf S}_{1 , \lambda}$ ($\lambda = 1,2,3$), which is
 given by

\beq
H_{B} = J_K  \{ S_{1,1}^x S_{2,1}^x + S_{2,1}^x S_{3,1}^x + S_{3,1}^x S_{1,1}^x\} + \delta H_B
\:\:\:\: ,
\label{yj.2}
\eneq
\noindent
with  ${\bf S}_{\lambda  , p }$ being the effective spin describing rhombus $p$ on
chain $\lambda$,  $J_K = \frac{{\cal J}^2}{J}$, and
$\delta H_B = - \frac{\sqrt{2}}{16} \frac{{\cal J}^2}{J}  \{ S_{1,1}^z + S_{2,1}^z + S_{3,1}^z \}$ is a
boundary magnetic field accounting for the modifications of the boundary conditions at the end-points of the
three chains forming the Y-network and affecting only the magnetic flux through the central region.

\noindent
As a
result, the Y-junction of rhombi chains is effectively described by the quantum spin Hamiltonian
$H_Y$, given by

\beq
H_{Y} = \sum_{ \lambda = 1,2,3} \{
J_x \sum_{ p = 1}^{\ell - 1}  S_{p , \lambda}^x S^x_{p+1 , \lambda} - 2  \sum_{p = 1}^\ell H_p  S_{p , \lambda}^z
\}+ H_{B}
\:\:\:\: .
\label{esham.5a}
\eneq
\noindent
$H_Y$ in Eq.(\ref{esham.5a}) is exactly Tsvelik's Hamiltonian for the Y-junction
of quantum spin chains (QSCJ) \cite{tse2}. When the 1QIMs are
driven near by the critical point ($J_x \sim 4 H$), the QSCJ model
in Eq.(\ref{esham.5a}) describes the two-channel Kondo model since the central region of the junction may be regarded as the 
effective protected spin-1/2 spin impurity discussed in \cite{tse2}.

If $E_C \gg J$ and $J \gg {\cal T}$ (this is a necessary
condition to safely rely on the description of each rhombus as an effective
spin-1/2 degree of freedom) one sees that the
condition $| 4 H | < J_x$, necessary to achieve the broken $Z_2$-symmetry phase in the
1QIM, is always satisfied; as a result \cite{tse2}, the 2CK-effect emerges, provided that the Kondo temperature
$T_K >   J_x -  4 | H|$ .

 Due to the correspondence between the microscopic
parameters of the JJRC and the macroscopic parameters of the 1QIMs, it is possible
to tune at will the parameters of the spin model by pertinently acting on the fabrication and control
parameters of the JJN. By acting on the Y-network control parameters,
it is possible to tune each  quantum Ising chains
nearby criticality ($J_x \approx 4 | H |$) by  locally changing the flux $\varphi$ piercing each rhombus; this amounts to modify $H$ by an amount $\sim J ( \varphi - \pi)$.

\begin{figure}
\centering \includegraphics*[width=0.75\linewidth]{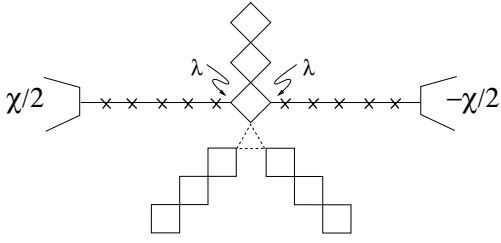}
\caption[]{Sketch of the circuit probing
$ m ( D )$: the rhombus at the endpoints of one chain is symmetrically
coupled to two one-dimensional Josephson junction arrays connected to
two bulk superconductors, whose phase difference is $\chi $.
At fixed $\chi$, the dc Josephson current across the one-dimensional Josephson arrays is
$\propto m ( D )$.  .}
\label{fig.3}
\end{figure}

{\it Probing the 2CK regime}
\\
To probe the 2CK regime emerging in a $Y$-junction of JJRCs one may use the circuit described 
in Fig.\ref{fig.3}. Namely, one may couple two opposite superconducting grains of a given rhombus to the 
endpoints of two one-dimensional quantum Josephson junction arrays (1JJA) coupled at their outer boundaries 
to two bulk superconductors set at a fixed phase difference $\chi$. We shall show in the following that the 
dc Josephson current flowing in the 1JJA as a result of this phase difference may be used to monitor the 
emergence of a 2CK regime in the $Y$ junction of JJRCs.

Conventional wisdom \cite{tse1} asserts that he onset of a Kondo regime is associated to scaling of 
physical observables with respect to a parameter , say $D$, typically chosen with the dimension of an 
energy (i.e.,  $D \sim  T $, or $D \sim J_x / \ell$). This happens for instance, to the magnetization 
next to the $Y$ junction defined as $m(D) = \langle S_{1 , \lambda}^z \rangle $ with $\lambda$ taken 
to be equal to 1 or 2 or 3. It is the behavior of $m(D)$ which can be monitored through the measurement 
of the dc-Josephson current flowing in the 1JJA. Indeed, the approach used in \cite{glark,giuso3} 
leads, after a somewhat tedious computation, to

\beq
I [ \chi ] = \frac{3}{2} \frac{\lambda^2}{J} \: m (D )  \:  \sin ( \chi )
\:\:\:\: ,
\label{cpa.3}
\eneq
\noindent
with $\lambda$ being the Josephson coupling between the endpoint of either
1JJA and the grain of the rhombus to which it is connected (see Fig.\ref{fig.3}.)
Eq.(\ref{cpa.3}) shows that to probe $m ( D )$ it is sufficient to monitor- at fixed $\chi$- $ I [ \chi ]$
for different values of $D$.

\begin{figure}
\centering \includegraphics*[width=0.9\linewidth]{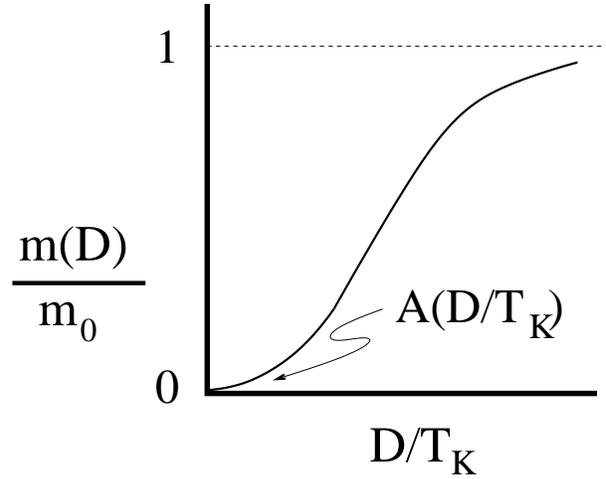}
\caption[]{$m ( D ) $ as a function of
$D / T_K$: the crossover from the perturbative behavior
 to the linear dependence on $D / T_K$ for $ D / t_K \ll 1$ is evidenced.}
\label{rhomb_6}
\end{figure}

The expected dependence of $m ( D ) $ on $D$ can be then inferred from
the standard analysis of the 2CK-problem \cite{tse1}. In particular, one
expects that the plot of   $ m ( D )  $ {\it vs.} $D/T_K$ takes the
form reported in Fig.\ref{rhomb_6}:
for $D / T_K \gg 1$, $m ( D ) $  starts from $m_0$ and decreases with a perturbative
correction $\propto J_K^2$, which logarithmically increases with $D$ as the cutoff approaches
 $T_K$. Eventually \cite{tse1}, the diagram turns into a linear dependence
of $m ( D )  $ on $D / T_K$ (which is a fingerprint of the 2CK-effect \cite{aflud,tse1,GiuTa2}),
as $ D \to 0$, finally flowing to 0 at the 2CK-fixed point.

{\it Concluding Remarks}
\\
In this paper we showed that a $Y$ junction of JJRCs may be used to simulate the two 
channel Kondo model recently proposed by Tsvelik in \cite{tse2}; in addition, we elucidated how 
the onset of the 2CK regime may be monitored through the measurement of a dc-Josephson current 
flowing in a 1JJA with a rhombus shaped impurity at its center. In our analysis we assumed that 
all the JJNs are fabricated with quantum junctions (i.e., with junctions such that the capacitive 
energy is much bigger than the Josephson energy) since, for these networks, it is much easier to 
exhibit the correspondence with spin models. However, this assumption is not crucial for our final 
results since a 1QIM may be realized also with networks fabricated with classical junctions 
\cite{ioffe_a}; using classical junctions has the great advantage of allowing to realize JJNs 
which are not only robust against the $1/f$ noise induced by stray charges in the array 
\cite{ioffe_a, propov} but also more accessible to direct measurements of current-phase characteristics \cite{pop}.

\noindent
\noindent

Acknowledgements: We benefited from discussions with A. Trombettoni, R. Egger, A. Ferraz, H. Johannesson, 
V. Korepin and A. Tagliacozzo. P. S. thanks CNPq for partial financial support through the grant 
provided by a {\it Bolsa de Produtivitade em Pesquisa}.

\end{document}